\documentstyle[12pt]{article}

\newcommand {\nn}    {\nonumber}
\newcommand {\vs}[1]  { \vspace*{#1 cm} }

\newcounter{eq}
\newcounter{sc}

\newcommand {\MPL}  {Mod. Phys. Lett.}
\newcommand {\IJMP}  {Int. J. Mod. Phys.}

\newcommand {\PL}   {Phys. Lett.}
\newcommand {\PR}   {Phys. Rev.}
\newcommand {\PRL}   {Phys. Rev. Lett.}

\newcommand {\PTP}  {Prog. Theor. Phys.}

\def\overleftrightarrow#1{\vbox{\ialign{##\crcr
 $\leftrightarrow$\crcr\noalign{\kern-1pt\nointerlineskip}
 $\hfil\displaystyle{#1}\hfil$\crcr}}}

\newlength{\minitwocolumn}
\begin{document}


\begin{flushright}
EDO-EP-46\\
July, 2003\\
\end{flushright}
\vspace{30pt}

\pagestyle{empty}
\baselineskip15pt

\begin{center}
{\large\bf Gravitational Localization of All Local Fields
on the Brane

 \vskip 1mm
}

\vspace{20mm}

Ichiro Oda
          \footnote{
          E-mail address:\ ioda@edogawa-u.ac.jp
                  }
\\
\vspace{10mm}
          Edogawa University,
          474 Komaki, Nagareyama City, Chiba 270-0198, JAPAN \\

\end{center}


\vspace{15mm}
\begin{abstract}
We present a new $(p - 1)$-brane solution to Einstein's equations in a general
space-time dimension. This solution is a natural generalization of the
stringlike defect solution with codimension 2 in 6 space-time dimensions, 
which has been recently discovered by Gogberashvili and Singleton, to a 
general $(p - 1)$-brane solution with codimension $n$ in general $D = p + n$
space-time dimensions.
It is shown that all the local fields are localized on the brane only
through the gravitational interaction although this solution does not
have a warp factor and takes a finite value in the radial infinity.
Thus, this solution is a solution in an arbitrary
space-time dimension realizing the idea of "gravitational trapping" 
of the whole bulk fields on the brane within the framework of a local 
field theory. Some problems associated with this solution and localization are
pointed out.

\vspace{15mm}

\end{abstract}

\newpage
\pagestyle{plain}
\pagenumbering{arabic}

\rm

The idea that our four dimensional world is a three-brane embedded 
in a higher dimensional space-time with non-factorizable warped geometry 
has been much investigated since the appearance of 
papers \cite{Randall1, Randall2, Gogberashvili1}.
(See also \cite{Akama, Rubakov, Visser} for the pioneering works and 
\cite{Oda1} for many brane model.)
In this idea, the key observation is that the graviton, which is allowed
to be free to propagate in the bulk, are confined to the brane
because of the warped geometry, thereby implying that the gravitational
law on the brane obeys the usual four dimensional Newton's law as
desired.

On the other hand, the other local fields except the gravitational field 
are not always localized on the brane even in the warped geometry.  Indeed, 
in the Randall-Sundrum model in five dimensions \cite{Randall2}, 
the following facts are well known: spin 0 field is localized
on a brane with positive tension which also localizes the graviton
\cite{Bajc}. Spin 1 field is not localized neither on a brane
with positive tension nor on a brane with negative tension
\cite{Pomarol}. (In six space-time dimensions, the spin 1 gauge field
is also localized on the brane \cite{Oda2}.) 
Moreover, spin 1/2 and 3/2 fields are localized not on a brane with positive 
tension but on a brane with negative tension \cite{Bajc}. 
Thus, in order to fulfill the localization of Standard Model particles 
on a brane with positive tension, it seems that some additional interactions 
except the gravitational interaction must be also introduced in the bulk. 
(See the review \cite{Oda3} and the papers \cite{Oda4} for the localization 
of the bulk fields in various brane world models.)

The introduction of such additional interactions, however, is not only
unnatural from the physical viewpoint but also can be applied to only  
specific situations. (For instance, for the localization of fermionic fields 
one must introduce a mass term with a 'kink' profile \cite{Jackiw}.) 
Thus, it is very welcoming if we could find a model in the brane world where 
all the local bulk fields are localized on the 3-brane only by the universal 
interaction, i.e., the gravity. It is of interest that Gogberashvili 
and Singleton \cite{Gogberashvili2, Gogberashvili3} have recently found such 
a solution to Einstein's equations in six space-time dimensions and pointed out 
that all the local fields ranging from the spin 0 scalar field to the spin 2 
gravitational field are localized on the 3-brane in this background geomery.

The aim of the present article is twofold. One aim is to extend their 3-brane 
model in six space-time dimensions to the case of a general $(p - 1)$-brane model 
in a general space-time dimension. We explicitly show that even in this general 
model whole local fields, those are, spin 0 scalar field, spin 1/2 spinor field, 
spin 1 gauge field, spin 3/2 gravitino field and spin 2 gravitational field 
as well as totally antisymmetric tensor fields, are confined on the $(p - 1)$-brane 
without appealing to the additional bulk interactions except the gravity.
The other aim is to mention some problems associated with the solution and
the localization. In particular, we will stress that the 'mild' localization of 
the wave function of the zero-modes might give rise to a conflict with experiments
and arguments about the stability of the brane is completely ignored.

The action which we consider in this article is that of gravity
in general $D = p + n$ dimensions, with the conventional Einstein-Hilbert
action plus the bulk cosmological constant and some matter action \cite{Vilenkin}:
\begin{eqnarray}
S = \frac{1}{2 \kappa_D^2} \int d^D x  
\sqrt{-g} \left(R - 2 \Lambda \right) 
+ \int d^D x  \sqrt{-g} L_m,
\label{1}
\end{eqnarray}
where $\kappa_D$ denotes the $D$-dimensional gravitational
constant with the relation $\kappa_D^2 = 8 \pi G_N = \frac{8 \pi}
{M_*^{D-2}}$ with $G_N$ and $M_*$ being the $D$-dimensional Newton 
constant and the $D$-dimensional Planck mass scale, respectively. 
Throughout this article we follow the standard conventions and notation 
of the textbook of Misner, Thorne and Wheeler \cite{Misner}. 

The variation of the action (\ref{1}) with respect to the $D$-dimensional
metric tensor $g_{MN}$ leads to Einstein's equations:
\begin{eqnarray}
R_{MN} - \frac{1}{2} g_{MN} R 
= - \Lambda g_{MN}  + \kappa_D^2 T_{MN},
\label{2}
\end{eqnarray}
where the energy-momentum tensor is defined as
\begin{eqnarray}
T_{MN} = - \frac{2}{\sqrt{-g}} \frac{\delta}{\delta g^{MN}}
\int d^D x \sqrt{-g} L_m.
\label{3}
\end{eqnarray}

We adopt the following metric ansatz:
\begin{eqnarray}
ds^2 &=& g_{MN} dx^M dx^N  \nn\\
&=& g_{\mu\nu} dx^\mu dx^\nu + \tilde{g}_{ab} dx^a dx^b  \nn\\
&=& \phi^2(r) \hat{g}_{\mu\nu}(x^\lambda) dx^\mu dx^\nu + g(r) (dr^2 
+ r^2 d \Omega_{n-1}^2),
\label{4}
\end{eqnarray}
where $M, N, ...$ denote $D$-dimensional space-time indices, 
$\mu, \nu, ...$ $p$-dimensional brane ones, and $a, b, ...$
$n$-dimensional extra spatial ones, so the equality $D=p+n$
holds. (We assume $p \ge 4$.) And d$\Omega_{n-1}^2$
stands for the metric on a unit $(n-1)$-sphere, which is 
concretely expressed in terms of the angular variables $\theta_i$ as
\begin{eqnarray}
d \Omega_{n-1}^2 = d\theta_2^2 + \sin^2 \theta_2 d\theta_3^2 
+ \sin^2 \theta_2 \sin^2 \theta_3 d\theta_4^2 + \cdots
+ \prod_{i=2}^{n-1} \sin^2 \theta_i d\theta_n^2,
\label{5}
\end{eqnarray}
with the volume element $\int d \Omega_{n-1} = \frac{2 \pi^{\frac{n}
{2}}}{\Gamma(\frac{n}{2})}$.

Moreover, following Gogberashvili and Singleton \cite{Gogberashvili2}
we take the ansatz for the energy-momentum tensor respecting the 
spherical symmetry (See \cite{Oda2} for the more general ansatz):
\begin{eqnarray}
T_{\mu\nu} &=& g_{\mu\nu} F(r),  \nn\\
T_{ab} &=& g_{ab} K(r),
\label{6}
\end{eqnarray}
where $F$ and $K$ are functions of only the radial coordinate $r$. 

Under these ansatze, after a straightforward calculation,
Einstein's equations reduce to
\begin{eqnarray}
&{}& \frac{1}{g} \Bigl[ p(p-1) \Bigl(\frac{\phi'}{\phi}\Bigr)^2 
+ p(n-1) \frac{\phi'}{\phi} \frac{(r^2 g)'}{r^2 g}
+ \frac{(n-1)(n-2)}{4} \Bigl(\frac{(r^2 g)'}{r^2 g}\Bigr)^2 \nn\\
&{}& - (n-1)(n-2) \frac{1}{r^2} \Bigr] 
- \frac{1}{\phi^2} \hat{R} + 2\Lambda = 2 \kappa_D^2 K,
\label{7}
\end{eqnarray}
\begin{eqnarray}
&{}& \frac{1}{g} \Bigl[ (p-1) \Bigl( 2 \frac{\phi''}{\phi} 
-\frac{g'}{g} \frac{\phi'}{\phi} \Bigr) 
+ (p-1)(p-2) \Bigl(\frac{\phi'}{\phi}\Bigr)^2 
+ (p-1)(n-1) \frac{\phi'}{\phi} \frac{(r^2 g)'}{r^2 g} \nn\\
&{}& + (n-1) \Big\{ \frac{(r^2 g)''}{r^2 g} + \frac{n-4}{4} 
\Bigl(\frac{(r^2 g)'}{r^2 g}\Bigr)^2 
- \frac{1}{2} \frac{g'}{g} \frac{(r^2 g)'}{r^2 g} 
- (n-2) \frac{1}{r^2} \Bigr\} \Bigr] \nn\\
&{}& + \frac{2-p}{p} \frac{1}{\phi^2} \hat{R} + 2\Lambda = 2 \kappa_D^2 F,
\label{8}
\end{eqnarray}
\begin{eqnarray}
&{}& \frac{1}{g} \Bigl[ p \Bigl( 2 \frac{\phi''}{\phi} 
- \frac{g'}{g} \frac{\phi'}{\phi} \Bigr) 
+ p(p-1) \Bigl(\frac{\phi'}{\phi}\Bigr)^2 
+ p (n-2) \frac{\phi'}{\phi} \frac{(r^2 g)'}{r^2 g} \nn\\
&{}& + (n-2) \Big\{ \frac{(r^2 g)''}{r^2 g} + \frac{n-5}{4} 
\Bigl(\frac{(r^2 g)'}{r^2 g}\Bigr)^2 
- \frac{1}{2} \frac{g'}{g} \frac{(r^2 g)'}{r^2 g} 
- (n-3) \frac{1}{r^2} \Bigr\} \Bigr] \nn\\
&{}& - \frac{1}{\phi^2} \hat{R} + 2\Lambda = 2 \kappa_D^2 K,
\label{9}
\end{eqnarray}
where the prime denotes the differentiation with respect to $r$,
and $\hat{R}$ is the scalar curvature associated with the
brane metric $\hat{g}_{\mu\nu}$.
Here we define the cosmological constant $\Lambda_p$ on the 
$(p-1)$-brane by the equation
\begin{eqnarray}
\hat{R}_{\mu\nu} - \frac{1}{2} \hat{g}_{\mu\nu} \hat{R} 
= - \Lambda_p \hat{g}_{\mu\nu}.
\label{10}
\end{eqnarray}
In deriving Eq. (\ref{8}), we have used  
$\hat{R}_{\mu\nu} = \frac{1}{p} \hat{g}_{\mu\nu} \hat{R}$, which
is obtained by taking the trace of Eq. (\ref{10}). Note that 
since $T_{\mu\nu}$ is proportional to $\hat{g}_{\mu\nu}$, 
$\hat{R}$ is a constant \cite{Vilenkin}.
In addition, the conservation law for the energy-momentum tensor,
$\nabla^M T_{MN} = 0$, takes the form
\begin{eqnarray}
K' + p \frac{\phi'}{\phi} (K - F) = 0.
\label{11}
\end{eqnarray}

One of our purposes in this article is to find a new $(p - 1)$-brane solution 
to Einstein's equations and the conservation law in the above. To do
that, the first step is to subtract Eq. (\ref{7}) from Eq. (\ref{9}). 
The result is given by
\begin{eqnarray}
2p \Bigl[ \frac{\phi''}{\phi} - \frac{g'}{g} \frac{\phi'}{\phi} 
- \frac{\phi'}{r \phi} \Bigr] + (n-2) \Big[ \frac{g''}{g} 
- \frac{3}{2} \Bigl(\frac{g'}{g}\Bigr)^2 
- \frac{1}{r} \frac{g'}{g} \Bigr] = 0.
\label{12}
\end{eqnarray}
Next we require the terms in each square bracket to vanish separately, 
that is,
\begin{eqnarray}
\frac{\phi''}{\phi} - \frac{g'}{g} \frac{\phi'}{\phi} 
- \frac{\phi'}{r \phi} &=& 0, \nn\\
\frac{g''}{g} - \frac{3}{2} \Bigl(\frac{g'}{g}\Bigr)^2 
- \frac{1}{r} \frac{g'}{g} &=& 0.
\label{13}
\end{eqnarray}
Here we notice that $n = 2$ is special in that the latter equation
does not arise from Eq. (\ref{12}) owing to the factor $n-2$.
In this sense, the stringlike defect solution with codimension 2,
which has been found by Gogberashvili and Singleton \cite{Gogberashvili2}
is distinct from the other defect solutions. Nevertheless, we will
see that there is a similar solution even in $n \neq 2$, whose solution
precisely corresponds to $b = 2$ case in \cite{Gogberashvili2}.

It turns out that the solution to the former equation in Eq. (\ref{13}) 
is given by \cite{Gogberashvili2}
\begin{eqnarray}
g(r) = \rho^2 \frac{\phi'(r)}{r},
\label{14}
\end{eqnarray}
where $\rho$ is an integration constant. The latter equation
in Eq. (\ref{13}) is then solved and the explicit forms of $\phi$ and $g$
are given by 
\begin{eqnarray}
\phi(r) &=& a \frac{r^2 - c^2}{r^2 + c^2}, \nn\\
g(r) &=& 4 a c^2 \rho^2 \frac{1}{(r^2 + c^2)^2},
\label{15}
\end{eqnarray}
where $a$, $c$ and $\rho$ are integration constants. (See below about boundary
conditions which we take.) Furthermore, we can show that the remaining Einstein's 
equations and the conservation law of the energy-momentum tensor are 
satisfied if we choose the following form of the source functions:
\begin{eqnarray}
K(r) &=& \frac{1}{2 \kappa_D^2} \Bigl[ \frac{4 c^2}{a \rho^2} 
p (p - 1) \frac{r^2}{(r^2 - c^2)^2} - \frac{1}{a \rho^2} 
(n - 1)(n + 2p -2) - \frac{1}{a^2} \Bigl(\frac{r^2 + c^2}{r^2 - c^2} \Bigr)^2 
\hat{R} + 2 \Lambda \Bigr], \nn\\
F(r) &=& \frac{1}{2 \kappa_D^2} \Bigl[ \frac{4 c^2}{a \rho^2} 
p (p - 1) \frac{r^2}{(r^2 - c^2)^2} - \frac{1}{a \rho^2} (n - 1)(n + 2p -2)
- \frac{2}{a \rho^2} (p - 1) \Bigl(\frac{r^2 + c^2}{r^2 - c^2} \Bigr)^2  \nn\\ 
&+& \frac{1}{a^2} \frac{2-p}{p} \Bigl(\frac{r^2 + c^2}{r^2 - c^2}\Bigr)^2 
\hat{R} + 2 \Lambda \Bigr].
\label{16}
\end{eqnarray}
Note that these source functions approach a definite value at the infinity 
$r \rightarrow \infty$. Although we could take $K(\infty) = F(\infty) = 0$
by selecting both $\hat{R}$ and $\rho$ (or $a$) appropriately, 
we do not so since $\phi$ in the solution (\ref{15}) does not take
the vanishing value at $r \rightarrow \infty$, either.

Here we should mention one subtle point, that is, what boundary conditions 
on the brane (and/or at the infinity) we should impose. 
For instance, in the previous work of the stringlike defect model with
codimension 2 \cite{Oda2}, we have required that the extra two 
dimensions are conical around the brane with a deficit angle in
order to describe the "local cosmic string" sitting at the origin
$r = 0$. In the case at hand, we take the different boundary
conditions which require us only to avoid singularities on the brane
\cite{Gogberashvili2}. Then, the suitable boundary conditions which we
take in this article are
\begin{eqnarray}
\phi(\varepsilon) &=& 1, \nn\\
\phi(\infty) &=& a,
\label{17}
\end{eqnarray}
where $\varepsilon$ denotes the "brane width", which now takes a fixed
value. The former boundary condition allows us to express the constant 
$c$ in terms of the "brane width" as $c = \sqrt{\frac{a - 1}{a + 1}} 
\varepsilon$, which implies $a > 1$ under the assumption of $a$ being 
positive. Let us count the number of independent integration constants 
in the solution (\ref{15}). Originally we have two second-order differential 
equations with respect to $\phi$ and $g$. Since we have set up two boundary 
conditions (\ref{17}), the number of the remaining independent constants 
should be two, which are nothing but $a$ and
$\rho$. Furthermore, it is worthwhile to mention that in this article 
the brane is assumed to have the nonvanishing "brane width" since 
the "brane width" $\varepsilon$ appears in the later arguments of localization 
of the bulk fields and plays a role as the  short-distance cutoff. 
In this context, let us note that the core physics 
inside $r = \varepsilon$ is in essence controlled by the short-distance and
high-energy physics, so the complete understanding of the core physics 
calls for quantum gravity. Because we have at present no satisfying theory 
of quantum gravity, it is physically reasonable to introduce such a cutoff 
via the boundary conditions into our theory where the cutoff has the physical
meaning as the brane width.   

Closely relating to the problem of boundary conditions, it is worthwhile to 
see how our solution (\ref{15}) can be described in the coordinate system
where the defects with a deficit angle are usually described. The result is
given by 
\begin{eqnarray}
ds^2 &=& \varphi^2(R) \hat{g}_{\mu\nu} dx^\mu dx^\nu + dR^2 
+ h(R) d \Omega_{n-1}^2 \nn\\
&=& a^2 \cos^2 \Bigl(\frac{1}{\rho \sqrt{a}} R \Bigr) \hat{g}_{\mu\nu} dx^\mu dx^\nu 
+ dR^2 + 4 a \rho^2 \sin^2 \Bigl(\frac{1}{2 \rho \sqrt{a}} R \Bigr) d \Omega_{n-1}^2. 
\label{18}
\end{eqnarray}
In this coordinate system, the line element has especially a simple
form in that the scale factor is expressed by (the square of) the cosine 
whereas the angular factor is the sine. Then, a deficit angle can be
calculated to $\delta = 2 \pi \frac{\varepsilon^2}{8 \rho^3 a}$,
which means that we have no deficit angle around the brane when
$\varepsilon \approx 0$ as expected \cite{Oda2}. 

Now we turn our attention to the problem of the localization of 
the bulk fields on the brane in the background geometry (\ref{15}).
Of course, in due analysis, we will neglect the back-reaction
on the geometry induced by the existence of the bulk fields.  
We proceed our study of the localization in order according to the size 
of spin of local fields and finally investigate totally antisymmetric tensor 
fields. 

Let us start with a massless, spin 0, real scalar coupled to gravity:
\begin{eqnarray}
S_0 = - \frac{1}{2} \int d^D x \sqrt{-g} g^{M N}
\partial_M \Phi \partial_N \Phi,
\label{19}
\end{eqnarray}
{}from which the equation of motion can be derived:
\begin{eqnarray}
\frac{1}{\sqrt{-g}} \partial_M (\sqrt{-g} g^{M N} \partial_N \Phi) = 0.
\label{20}
\end{eqnarray}
{}From now on, without loss of generality, we shall take 
a flat metric on the brane, that is, $\hat{g}_{\mu\nu} = \eta_{\mu\nu}$.
It turns out that $\Phi(x^M) = \phi(x^\mu) u_0$ which satisfies 
the Klein-Gordon equation on the brane 
$\eta^{\mu\nu} \partial_\mu \partial_\nu \phi(x) = 0$ is a solution to 
the equation of motion (\ref{20}) in the background metric (\ref{15}).
Substituting this solution into the starting action (\ref{19}), the
action can be cast to
\begin{eqnarray}
S_0 = - \frac{1}{2} \frac{2 \pi^{\frac{n}{2}}}{\Gamma(\frac{n}{2})} u_0^2 
\int_\varepsilon^{\infty} dr \phi^{p-2} g^{\frac{n}{2}} r^{n-1} 
\int d^p x \eta^{\mu\nu} \partial_\mu \phi \partial_\nu \phi
+ \cdots,
\label{21}
\end{eqnarray}

Now we wish to show that this zero-mode is localized on the
brane sitting around the origin $r=0$. The condition for
having localized $p$-dimensional scalar field is that the solution
is normalizable. It is of importance to notice that normalizability
of the ground state wave function is equivalent to the condition that the 
"coupling" constant is nonvanishing. In other words, in order to show
that the bulk zero-modes which satisfy the equation of motion in the
bulk is in fact confined to a brane, the zero-modes must give us the kinetic
terms on the brane, from which we can understand that the bulk zero-modes
are truely dynamical and propagate on the brane. Thus, provided that 
we define
\begin{eqnarray}
I_0 &=& \int_\varepsilon^{\infty} dr \phi^{p-2} g^{\frac{n}{2}} r^{n-1} \nn\\
&=& (2 c \rho)^n a^{p + \frac{n}{2} -2} \int_\varepsilon^{\infty} dr 
\frac{(r^2 - c^2)^{p-2}}{(r^2 + c^2)^{n+p-2}}  r^{n-1},
\label{22}
\end{eqnarray}
the condition of having localized $p$-dimensional scalar field on the
brane requires that $I_0$ should be finite. The integrand in $I_0$
scales as $\frac{1}{r^{n+1}}$ at the radial infinity and is a smooth
function between $r = \varepsilon$ and $r = \infty$, so $I_0$ is finite
even if the analytic expression is not available. (In the case of
$\varepsilon = c$, $I_0$ can be expressed in terms of the hypergeometric
function, which is of course finite.) Hence, the $p$-dimensional 
scalar field $\phi$ is localized on the brane by the gravitational
interaction.

Next, let us consider spin 1/2 spinor field.
Our starting action in this case is the Dirac action given by
\begin{eqnarray}
S_{\frac{1}{2}} = \int d^D x \sqrt{-g} \bar{\Psi} i \Gamma^M D_M \Psi,
\label{23}
\end{eqnarray}
{}from which the equation of motion is given by
\begin{eqnarray}
0 = \Gamma^M D_M \Psi = (\Gamma^\mu D_\mu + \Gamma^r D_r +
\Gamma^{\theta_i} D_{\theta_i}) \Psi.
\label{24}
\end{eqnarray}

We introduce the vielbein $e_M ^{\bar{M}}$ (and its inverse 
$e^M _{\bar{M}}$) through the usual definition $g_{M N} = e_M ^{\bar{M}} 
e_N ^{\bar{N}} \eta_{\bar{M} \bar{N}}$ where $\bar{M}, \bar{N}, \cdots$ 
denote the local Lorentz indices. $\Gamma^M$ in a curved space-time
is related to $\gamma^{\bar{M}}$ in a flat space-time by 
$\Gamma^M = e^M _{\bar{M}} \gamma^{\bar{M}}$.
In addition, the spin connection $\omega_M^{\bar{M} \bar{N}}$ in the covariant 
derivative $D_M \Psi = (\partial_M + \frac{1}{4} \omega_M^{\bar{M} 
\bar{N}} \gamma_{\bar{M} \bar{N}}) \Psi \equiv (\partial_M + \omega_M) \Psi$ 
is defined as
\begin{eqnarray}
\omega_M ^{\bar{M} \bar{N}} &=& \frac{1}{2} e^{N \bar{M}} 
(\partial_M e_N ^{\bar{N}} - \partial_N e_M ^{\bar{N}})
- \frac{1}{2} e^{N \bar{N}} 
(\partial_M e_N ^{\bar{M}} - \partial_N e_M ^{\bar{M}}) \nn\\
&-& \frac{1}{2} e^{P \bar{M}} e^{Q \bar{N}}
(\partial_P e_{Q {\bar{R}}} - \partial_Q e_{P {\bar{R}}})
e^{\bar{R}} _M,
\label{25}
\end{eqnarray}
so the covariant derivative can be calculated to
\begin{eqnarray}
D_\mu \Psi &=& (\partial_\mu +  \frac{1}{2} \frac{\phi'}{g \phi} 
\Gamma_\mu \Gamma_r) \Psi, \nn\\
D_r \Psi &=& \partial_r \Psi, \nn\\  
D_{\theta_i} \Psi &=& \Bigl[ \partial_{\theta_i} - \frac{1}{2} 
\frac{1}{g^{\frac{3}{2}} r} \partial_r (g^{\frac{1}{2}} r) \Gamma_r
\Gamma_{\theta_i} + \tilde{\omega}_{\theta_i}(\theta) \Bigr] \Psi,
\label{26}
\end{eqnarray}
where $\tilde{\omega}_{\theta_i}(\theta)$ is a contribution from
$S^{n-1}$, whose explicit form is now irrelevant so is omitted to
write down.

Let us look for a solution with the form of $\Psi(x^M) = \psi(x^\mu) 
u(r) \chi(\theta)$, where $\psi(x^\mu)$ satisfies the massless 
$p$-dimensional Dirac equation $\gamma^\mu \partial_\mu \psi = 0$ and 
the chiral condition $\gamma^r \psi(x^\mu) = \psi(x^\mu)$, and 
$\chi$ satisfies the equation $\gamma^{\theta_i} (\partial_{\theta_i}
+ \tilde{\omega}_{\theta_i}) \chi = 0$. With this ansatz, the Dirac
equation (\ref{24}) is reduced to
\begin{eqnarray}
\Bigl[ \partial_r +  \frac{p}{2} \frac{\phi'}{\phi} + \frac{n-1}{2} 
\frac{\partial_r(g^{\frac{1}{2}} r)} {g^{\frac{1}{2}} r} \Bigr]
u(r) = 0.
\label{27}
\end{eqnarray}
The solution to this equation then reads:
\begin{eqnarray}
u(r) = c_{\frac{1}{2}} \phi^{- \frac{p}{2}} 
(g^{\frac{1}{2}} r)^{- \frac{n-1}{2}},
\label{28}
\end{eqnarray}
with $c_{\frac{1}{2}}$ being an integration constant. 

Now we are willing to show that the solution (\ref{28}) is normalizable 
so that the spin 1/2 spinor field is localized on the brane. Inserting
the above solution to the action gives rise to
\begin{eqnarray}
S_{\frac{1}{2}} = \int_\varepsilon^{\infty} dr \phi^{p-1} g^{\frac{n}{2}}
r^{n-1} u^2(r) \int d \Omega_{n-1} \chi^2(\theta) 
\int d^p x \bar{\psi} i \gamma^\mu \partial_\mu  \psi + \cdots.
\label{29}
\end{eqnarray}
In order to localize the spin 1/2 fermion, the integral $I_{\frac{1}{2}}$, 
which is defined as
\begin{eqnarray}
I_{\frac{1}{2}} = \int_\varepsilon^{\infty} dr \phi^{p-1} g^{\frac{n}{2}}
r^{n-1} u^2(r)
\label{30}
\end{eqnarray}
should be finite. (Here note that the integral over $S^{n-1}$ is 
finite.)  Indeed, this integral can be easily calculated as
\begin{eqnarray}
I_{\frac{1}{2}} = \frac{c_{\frac{1}{2}}^2 \rho}{\sqrt{a}} 
\log \Bigl|\frac{\varepsilon
+ c}{\varepsilon - c} \Bigr|,
\label{31}
\end{eqnarray}
which is obviously finite as long as the brane width $\varepsilon$
is nonvanishing. The important point here is that the divergence
at $r = \infty$, which usually makes the zero-mode solutions 
unnormalizable, does not occur in the model at hand.

Let us turn to spin 1 gauge field. We consider the action
of $U(1)$ vector field:
\begin{eqnarray}
S_1 = - \frac{1}{4} \int d^D x \sqrt{-g} g^{M N} g^{R S}
F_{MR} F_{NS},
\label{32}
\end{eqnarray}
where $F_{MN} = \partial_M A_N - \partial_N A_M$ as usual.
{}From this action the equation of motion is given by
\begin{eqnarray}
\frac{1}{\sqrt{-g}} \partial_M (\sqrt{-g} g^{M N} g^{R S} F_{NS}) = 0.
\label{33}
\end{eqnarray}
It is easily checked that $A_{\mu}(x^M) 
= a_\mu(x^\lambda) u_0$, $A_r(x^M)=$constant, and $A_{\theta_i}(x^M)=$constant
is a solution to this equation of motion if $\partial^\mu f_{\mu\nu} = 0$
where $f_{\mu\nu} \equiv \partial_\mu a_\nu - \partial_\nu a_\mu$.

Once we have found the solution, let us ask ourselves whether this solution 
is a normalizable one or not by substituting it into 
the action (\ref{32}). It turns out that the action is reduced to
\begin{eqnarray}
S_1 = - \frac{1}{4} \frac{2 \pi^{\frac{n}{2}}}{\Gamma(\frac{n}{2})} u_0^2 
\int_\varepsilon^{\infty} dr \phi^{p-4} g^{\frac{n}{2}} r^{n-1} 
\int d^p x \eta^{\mu\nu} \eta^{\lambda\sigma} 
f_{\mu\lambda} f_{\nu\sigma} + \cdots. 
\label{34}
\end{eqnarray}
The integral defined by 
\begin{eqnarray}
I_1 &=& \int_\varepsilon^{\infty} dr \phi^{p-4} g^{\frac{n}{2}} r^{n-1} 
\nn\\
&=& (2 c \rho)^n a^{p + \frac{n}{2} -4} \int_\varepsilon^{\infty} dr 
\frac{(r^2 - c^2)^{p-4} r^{n-1}}{(r^2 + c^2)^{n + p-4}} 
\label{35}
\end{eqnarray}
is finite as in the scalar field. Thus, the vector field can be also 
localized on the brane only by the gravitational interaction.

Next we are ready to consider spin 3/2 fermionic field, in other words, 
the gravitino.
Let us begin with the action of the Rarita-Schwinger gravitino
field:
\begin{eqnarray}
S_{\frac{3}{2}} = \int d^D x \sqrt{-g} \bar{\Psi}_M i \Gamma^{[M} \Gamma^N
\Gamma^{R]} D_N \Psi_R,
\label{36}
\end{eqnarray}
{}from which the equation of motion is given by
\begin{eqnarray}
\Gamma^{[M} \Gamma^N \Gamma^{R]} D_N \Psi_R = 0.
\label{37}
\end{eqnarray}
Here the square bracket denotes the total antisymmetrization and the
covariant derivative is defined with the affine connection 
$\Gamma^R_{MN} = e^R_{\bar{M}}(\partial_M e_N^{\bar{M}} +
\omega_M^{\bar{M} \bar{N}} e_{N {\bar{N}}})$ by
$D_M \Psi_N = \partial_M \Psi_N - \Gamma^R_{MN} \Psi_R 
+ \frac{1}{4} \omega_M^{\bar{M} \bar{N}} \gamma_{\bar{M} \bar{N}} 
\Psi_N$. We look for a solution with the form of
$\Psi_\mu(x^M) = \psi_\mu(x^\lambda) u(r) \chi(\theta)$ and
$\Psi_r = 0 = \Psi_{\theta_i}$ where $\psi$ and $\chi$ are
assumed to satisfy the equations $\gamma^\mu \psi_\mu 
= \gamma^{[\mu} \gamma^\nu \gamma^{\rho]}
\partial_\nu \psi_\rho = 0$, $\gamma^r \psi_\mu = \psi_\mu$ 
and $\gamma^{\theta_i} (\partial_{\theta_i}
+ \tilde{\omega}_{\theta_i}) \chi = 0$. Then, the equation of
motion (\ref{37}) takes the form
\begin{eqnarray}
\Bigl[ \partial_r +  \frac{p-1}{2} \frac{\phi'}{\phi} + \frac{n-1}{2} 
\frac{\partial_r(g^{\frac{1}{2}} r)} {g^{\frac{1}{2}} r} \Bigr]
u(r) = 0.
\label{38}
\end{eqnarray}
The solution to this equation reads:
\begin{eqnarray}
u(r) = c_{\frac{3}{2}} \phi^{- \frac{p-1}{2}} 
(g^{\frac{1}{2}} r)^{- \frac{n-1}{2}},
\label{39}
\end{eqnarray}
with $c_{\frac{3}{2}}$ being an integration constant. 

We shall show that as in the case of spin 1/2 field,
this solution is normalizable so the gravitino field is
also localized on the brane. To do so, let us substitute the solution
into the action, whose result is of form
\begin{eqnarray}
S_{\frac{3}{2}} = \int_\varepsilon^{\infty} dr 
\phi^{p-3} g^{\frac{n}{2}}r^{n-1} u^2(r)  
\int d \Omega_{n-1} \chi^2(\theta) \int d^p x \bar{\psi}_\mu i \gamma^{[\mu} 
\gamma^\nu \gamma^{\rho]} \partial_\nu \psi_\rho + \cdots.
\label{40}
\end{eqnarray}
It is certain that the integral $I_{\frac{3}{2}}$, which is defined as
\begin{eqnarray}
I_{\frac{3}{2}} &=& \int_\varepsilon^{\infty} dr \phi^{p-3} g^{\frac{n}{2}}
r^{n-1} u^2(r) \nn\\
&=& \frac{c_{\frac{3}{2}}^2 c \rho}{a \sqrt{a}} \frac{2 \varepsilon}{\varepsilon^2 - c^2}
\label{41}
\end{eqnarray}
is finite as long as the brane width $\varepsilon$ is non-zero.
(Here note that the integral over $S^{n-1}$ is also finite.) 

Next let us consider spin 2 gravitational field. As in the cases 
treated so far, we can search for a solution to the equation of motion
in the background (\ref{15}), insert the solution in the 
Einstein-Hilbert action and then examine the finiteness of the radial integral. 
However, in this case, it is well known that the localization property of 
the graviton is the same as in the scalar field \cite{Oda2}, so we can conclude 
that the bulk graviton is trapped on the brane as in case of the scalar field.

Finally, we take account of totally antisymmetric tensor fields. In general,
the action of $k$-rank totally antisymmetric tensor field $A_k$ is of 
the form in the form notation
\begin{eqnarray}
S_k = - \frac{1}{2} \int F_{k+1} \wedge * F_{k+1},
\label{42}
\end{eqnarray}
where $F_{k+1} = d A_k$. The equation of motion is simply given by
\begin{eqnarray}
d \wedge * F_{k+1} = 0.
\label{43}
\end{eqnarray}
We can show that $A_{\mu_1 \mu_2 \cdots \mu_k} = a_{\mu_1 \mu_2 \cdots \mu_k}(x^\lambda)
u_0$ is a solution to this equation of motion if $d \wedge * f = 0$ where
$f = d a$. Substituting this solution in the action (\ref{42}) leads to
the following expression:
\begin{eqnarray}
S_k = I_k  \int f_{k+1} \wedge * f_{k+1} + \cdots,
\label{44}
\end{eqnarray}
where $I_k$ is defined as
\begin{eqnarray}
I_k &\propto& \int_\varepsilon^{\infty} dr \phi^{p-2-2k} g^{\frac{n}{2}} r^{n-1} 
\nn\\
&\propto& \int_\varepsilon^{\infty} dr 
\frac{(r^2 - c^2)^{p-2-2k} r^{n-1}}{(r^2 + c^2)^{n+p-2-2k}}. 
\label{45}
\end{eqnarray}
It is obvious that $I_k$ is finite so the totally antisymmetric tensor
fields are also localized on the brane by the gravitational interaction.

In conclusion, in this article, we have presented a new $(p - 1)$-brane
solution in an arbitrary space-time dimension. This solution
is a natural generalization of the 3-brane solution in six dimensions
recently discovered by Gogberashvili and Singleton \cite{Gogberashvili2}
to general $D$ space-time dimensions.
We have also clarified that the stringlike defect model with codimension
2 is specific due to the terms proportional to the factor $n - 2$ in
Einstein's equations. Moreover, we have presented a complete analysis
of localization of all bulk fields on a brane and showed that all
the bulk fields are trapped on the brane only via the gravitational
interaction. It is well known that in the warped geometry spin 1/2 and
3/2 fermionic fields are not trapped by the gravitational interaction
so it is necessary to introduce a nontrivial Higgs coupling, thereby 
generating a bulk mass term with a 'kink' profile and leads to the
localization of these fermionic fields on the brane \cite{Jackiw}. 
It is remarkable that in the present model, we do not have to include 
such an additional interaction for the localization of the fermionic fields.

At this stage, it is useful to ask why our solution gives rise to the localization
of all the bulk fields on the brane. The technical reason is very much simple. Namely,
the scale factor $\phi(r)$ has a property such that it approaches a 
definite value at the infinity and a smooth function without singularities
from the edge of the brane to the radial infinity. So the normalizability
of the ground state wave function, which is equivalent to
a finite integral over the radial coordinate $r$, is assured for all
the bulk fields. On the other hand, in the warped geometry, 
the integral over $r$ associated with the fermionic fields includes
$e^{+ c r}$ ($c > 0$) factors coming from $g^{\mu\nu}$ (and $e^\mu_{\bar{\mu}}$), 
{}for which the integral over $r$ diverges at $r \rightarrow \infty$ 
so it leads to the non-localization of these fermionic fields. 

As problems in this model, we should first recall one important point about
the localization. 
As stressed before \cite{Oda4}, the normalizable condition, in other words, 
the convergence of the integral over $r$, is usually thought to be 
a condition for the localization, but the story is not so simple. 
As in the locally localized gravity models (see the third paper in \cite{Oda4})
the present model provides us with an example such that the zero-mode solutions
of the bulk fields are normalizable, but their wave functions spread rather widely 
in the bulk owing to the lack of the warp factor. Thus, in order not to contradict 
with the strict experiments 
such as the charge conservation law, some parameters in our model must be chosen 
in a proper way. At present, we have no idea whether there is such a suitable choice 
of the parameters.

As the second problem, we wish to point out a problem associated with the source 
functions. In our model, the presence of a solution to Einstein's equations
heavily depends on the form of the source functions. Therefore, there would be
a possibility that we might have more solutions by changing the form of the
source functions. The real problem is then how to construct such source
functions from fundamental matter fields so that the brane is a stable
localized object. For instance, a set of $n$ scalar functions with the Higgs
potential, thereby breaking the $\it{global}$ $SO(n)$ symmetry to $SO(n-1)$
symmetry, are used to make the topologically $\it{stable}$ brane since
a topological argument guarantees the stability because of a mathematical
formula $\Pi_{n-1}(SO(n-1)) = Z$ \cite{Oda4}.

Let us close by mentioning some interesting future works related to the present study.
For instance, one interesting work would be to construct a supergravity model 
corresponding to the situation at hand and investigate the SUSY-breaking
and the cosmological constant problem e.t.c. The other problem
is to make the source functions concretely from some local
field such as the scalar field. As mentioned above, the physics near
the core of the brane is in the regime of the short-distance and
the high-energy physics, so it would be difficult to understand
the physics completely since it is expected that quantum gravity plays an 
important role in the core physics. However, some low energy effective action
might be useful to describe the characteristic behavior of our source 
functions and insure the stability of a brane under deformations. We wish to
report these works in future publication.

\begin{flushleft}
{\bf Acknowledgements}
\end{flushleft}

This work has been partially supported by the grant from 
the Japan Society for the Promotion of Science, No. 14540277.

\vs 1

\end{document}